\documentclass[prl,twocolumn,superscriptaddress,floatfix,noshowpacs,10pt,longbibliography]{revtex4-1}%

\usepackage{graphicx,bm,times}
\usepackage{amsmath}
\usepackage{amsfonts}
\usepackage{amssymb}

\usepackage{bm}% bold math
\usepackage{color}
\usepackage{times}

\usepackage[%
colorlinks=true,
urlcolor=blue,
linkcolor=blue,
citecolor=blue
]{hyperref}

\begin{document}

\title{Competing pairing interactions
	responsible for the large upper critical field \\ in a stoichiometric iron-based superconductor, CaKFe$_4$As$_4$ }

\author{Matthew Bristow}
\email[corresponding author: ]{matthew.bristow@physics.ox.ac.uk}
\affiliation{Clarendon Laboratory, Department of Physics,
University of Oxford, Parks Road, Oxford OX1 3PU, UK}

\author{William Knafo}
\affiliation{Laboratoire National des Champs Magn\'{e}tiques Intenses (LNCMI), CNRS-UJF-UPS-INSA, Toulouse, France}

\author{Pascal Reiss}
\affiliation{Clarendon Laboratory, Department of Physics,
	University of Oxford, Parks Road, Oxford OX1 3PU, UK}

 \author{William Meier}
 \affiliation{Ames Laboratory, Iowa State University, Ames, Iowa 50011, USA}
 \affiliation{Department of Physics and Astronomy, Iowa State University, Ames, Iowa 50011, USA}

  \author{Paul C. Canfield}
 \affiliation{Ames Laboratory, Iowa State University, Ames, Iowa 50011, USA}
 \affiliation{Department of Physics and Astronomy, Iowa State University, Ames, Iowa 50011, USA}

 \author{Stephen J. Blundell}
 \affiliation{Clarendon Laboratory, Department of Physics,
 	University of Oxford, Parks Road, Oxford OX1 3PU, UK}

\author{Amalia I. Coldea}
\email[corresponding author: ]{amalia.coldea@physics.ox.ac.uk}
\affiliation{Clarendon Laboratory, Department of Physics,
University of Oxford, Parks Road, Oxford OX1 3PU, UK}

\begin{abstract}
The upper critical field of multi-band superconductors is an important quantity that can reveal the details about the nature of the superconducting pairing.
Here we experimentally map out the complete upper critical field phase diagram of a stoichiometric superconductor, CaKFe$_{4}$As$_{4}$, up to $90\,$T  for different orientations of the magnetic field and at temperatures down to $4.2\,$K.
The upper critical fields are extremely large, reaching values close to $\sim3T_{\rm{c}}$ at the lowest temperature, and the anisotropy decreases dramatically with temperature leading to essentially isotropic superconductivity at $4.2\,$K.
We find that the temperature dependence of the upper critical field can be well described by a two-band model in the clean limit with band coupling parameters favouring intraband  over interband interactions.
The large Pauli paramagnetic effects together with the presence of the shallow bands is consistent with the stabilization of an FFLO state at low temperatures in this clean superconductor.
\end{abstract}
\date{\today}
\maketitle

The upper critical field, $H_{\rm{c2}}$, is an important property of superconductors
that defines their limit for practical applications.
It also describes the complex interplay between different  pairing gaps and symmetry and
 can shed light on
the nature of the superconducting mechanism.
Furthermore, the temperature dependence of the upper critical field can also
provide evidence for the presence of the Fulde-Ferrell-Larkin-Ovchinnikov (FFLO) state \cite{gurevich2010upper,Song2019}
in which the order parameter varies in space.
The iron-based superconductors have unusually
  large values of the upper critical field which
 reveal exotic effects caused by the interplay
of orbital and paramagnetic pair-breaking in multiband
superconductors with unconventional pairing symmetry \cite{Hirschfeld2011}.
They also provide the right conditions
for the FFLO state to develop in clean materials
due to the likely presence of shallow bands \cite{Mou2016} and very large
Pauli paramagnetic effects \cite{gurevich2010upper}.

CaKFe$_{4}$As$_{4}$  is a clean and stoichiometric superconductor with  a relatively high $T_{\rm{c}}=35\,$K
and it belongs to a new family of 1144 iron-based superconductors \cite{Iyo2016}.
This system lacks long-ranged magnetic order or a nematic electronic state at low temperatures
  \cite{Iyo2016,Iida2017,Meier2016,Bud2017,Bud2018,Cui2017,Zhang2018}
but upon doping with Ni a hedgehog magnetic structure is stabilized  \cite{Ding2017,Ding2018b,Bud2018}.
CaKFe$_{4}$As$_{4}$ has an exceptionally large critical current density
due to the strong point-like defects caused by local structural site effects as well as surface pinning \cite{Singh2018,Haberkorn2019,Ishida2019}.
Due to reduced symmetry compared with the 122 family of iron-based superconductors, CaKFe$_{4}$As$_{4}$
is predicted to have up to ten different bands (Fig. \ref{Fig1}(e) and Fig. \ref{FigSM:FS} in the Appendix).
However, angle resolved photoemission spectroscopy
detects a Fermi surface composed of three hole pockets and two electron pockets.
The superconducting gaps are nearly isotropic and different for each of the
Fermi surface sheets \cite{Mou2016}.
The presence of electron and hole sheets supports
a spin resonance corresponding to the ($\pi, \pi$)
nesting wave vector  detected by neutron diffraction \cite{Iida2017}
and promotes a $s_{\pm}$  superconducting pairing symmetry in CaKFe$_{4}$As$_{4}$.
Thus, this clean system is a model system for understanding the effect of
pairing on its upper critical field.

In order to understand the superconducting properties of CaKFe$_{4}$As$_{4}$
we have measured the upper critical fields for two orientations in magnetic fields up to $90\,$T using electrical transport measurements.
These experimental studies provide a complete $H_{\rm{c2}}(T)$ phase diagram and allow us to model the entire temperature dependence,
as previous work in magnetic fields up to $60\,$T could not reach the low temperature region
 \cite{Meier2016}.
We find that CaKFe$_{4}$As$_{4}$ is highly isotropic at the
lowest temperature. A two-band model describes the dependence of the upper critical fields
for both directions and the band coupling parameters
indicate the presence of different pairing channels.
At low temperatures, the upper critical field does not saturate but shows an upturn,
consistent with the emergence of a FFLO state in CaKFe$_{4}$As$_{4}$.

\begin{figure*}[htbp]
	\centering
	\includegraphics[trim={0cm 0cm 0cm 0cm}, width=1\linewidth,clip=true]{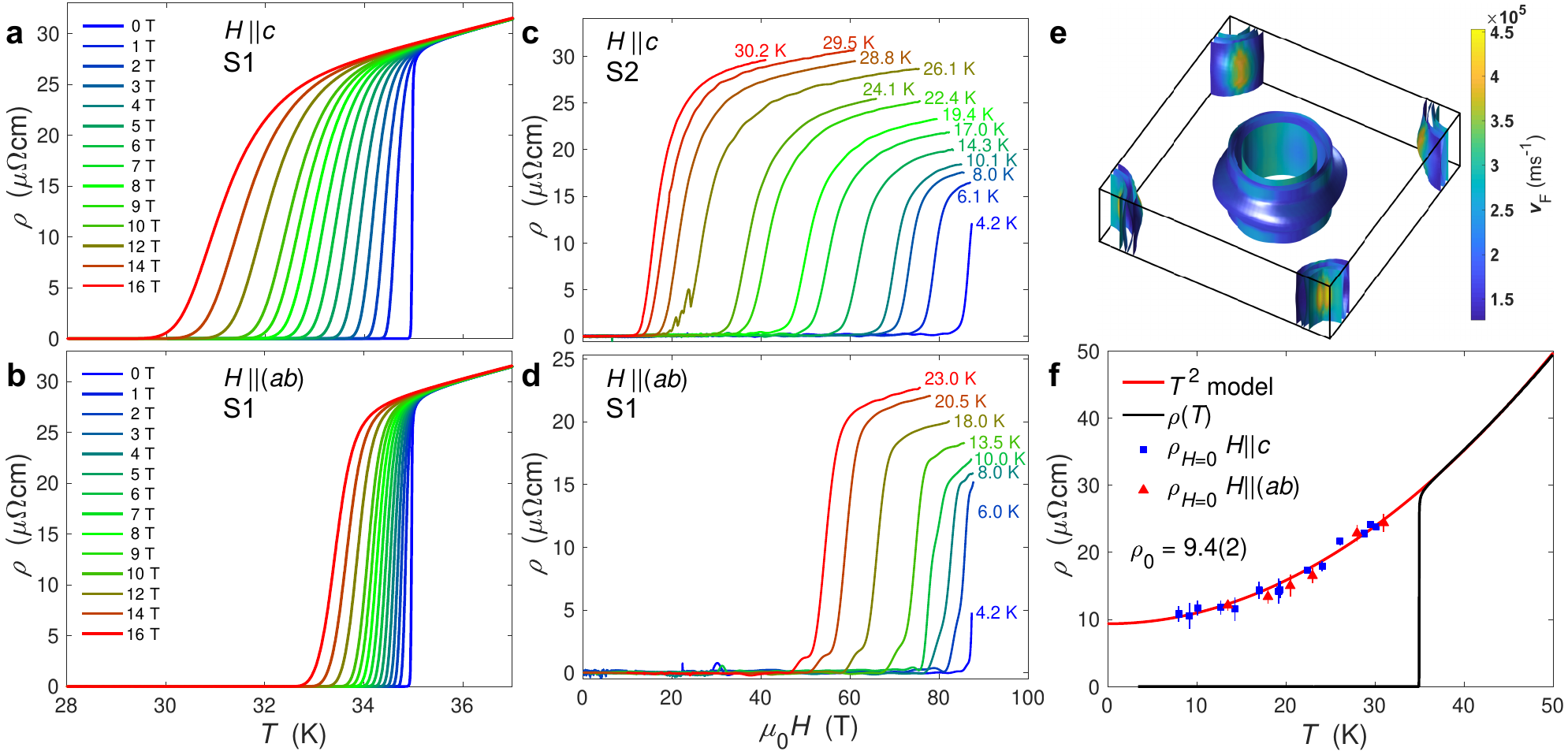}
	\caption{Resistivity versus temperature for CaKFe$_{4}$As$_{4}$ (S1)
 at different constant magnetic fields for (a) $H||c$ and (b) $H||(ab)$.
 Resistivity versus magnetic field at constant temperatures measured concomitantly
 in pulsed fields up $90\,$T, for (c) $H||c$ (S2) and (b) $H||(ab)$ (S1).
 (e) Fermi surface of CaKFe$_{4}$As$_{4}$  using experimental lattice parameters (as detailed in Fig.~\ref{FigSM:FS} in the Appendix).
 The colours reflect the variation of the Fermi velocity.
 (f) Resistivity against temperature for sample S1 together with the zero-field extrapolated values from high fields
 shown in (c) and (d) (see also Fig.~\ref{FigSM_rhoT}(d) in the Appendix).
The zero-temperature resistivity $\rho_{\rm{0}}$ is estimated as $9.4(2)\,\rm{\mu\Omega{}cm}$
using a Fermi-liquid $T^2$ behaviour (red line in (f)).
}
	\label{Fig1}
\end{figure*}

{\bf Experimental details.}
Single crystals of CaKFe$_{4}$As$_{4}$ were grown in the Ames Laboratory, as discussed elsewhere \cite{Meier2017,Iyo2016}.
Robust electrical contacts were achieved using indium solder, giving contact resistances of less than $0.5\,\mathrm{\Omega}$.
Samples were measured using the standard 4- and 5-point $ac$ lock-in techniques and a current of $1\,$mA.
We  have investigated several high quality single crystals, with large residual resistivity ratios  $\sim 14.5$
($RRR=\rho(300\rm{K})/\rho(36\rm{K}$)), small residual resistivity, $\rho_{\rm{0}} \sim9.4\,\mathrm{\mu\Omega{}cm}$,
and sharp superconducting transitions, $\Delta T_{\rm{c}} \sim 0.1\,$K (see Figs.~\ref{Fig1}(f) and \ref{FigSM_rhoT}(c) in the Appendix).
Transport measurements were performed in Oxford up to $16\,$T in a Quantum Design PPMS
in constant magnetic fields and for different magnetic field orientations ($H||c$ and $H||$($ab$)).
Pulsed magnetic field measurements were performed at LNCMI Toulouse,
using a 70~T single coil and a 90~T dual coil
at constant temperatures below $35\,$K for both field orientations.
Magnetic fields up to $90\,$T were produced by using current pulses through
two different solenoid coils (an example is shown in Fig.~\ref{FigSM_PulseExample} in the Appendix).

{\bf Upper critical field from transport measurements.}
Fig.~\ref{Fig1} shows resistivity against temperature for different fixed magnetic fields
for orientations of the sample in relation to the applied magnetic field.
As the magnetic field increases the superconducting transition becomes broader and suppressed faster
for $H||c$ by $5\,$K in $16\,$T (Fig.~\ref{Fig1}(a)),
as compared with the $H||(ab)$ case
for which $H_{\rm c2}$
 only changes by  $2\,$K in $16\,$T (Fig.~\ref{Fig1}(b)),
as the orbital effects are less effective in suppressing superconductivity in this orientation.
 These data are used to extract the upper critical field near $T_{\rm{c}}$,
 defined as the offset field, as shown in Fig.~\ref{FigSM_rhoT}(a) and (b) in the Appendix.
Due to the high crystallinity of our samples that display sharp superconducting transitions,
we find that the critical temperature from magnetization measurements is similar to the
offset temperature determined from transport measurements (Fig.~\ref{FigSM_rhoT}(a)).
  In order to completely suppress the superconductivity of CaKFe$_{4}$As$_{4}$
 we have used pulsed magnetic fields up to $90\,$T.
Resistivity data against magnetic fields up to $90\,$T measured at fixed temperature are shown in Fig.~\ref{Fig1}(c) and \ref{Fig1}(d)
for the two different orientations. Despite the strong disparity in the degree of suppression of superconductivity
between the two field directions close to $T_{\rm c}$,
at the lowest measurable temperature of $4.2\,$K
the normal state is reached at a similar field $\sim85\,$T for both orientations.

\begin{figure}[htbp]
	\centering
\includegraphics[trim={0cm 0cm 0cm 0cm}, width=0.80\linewidth,clip=true]{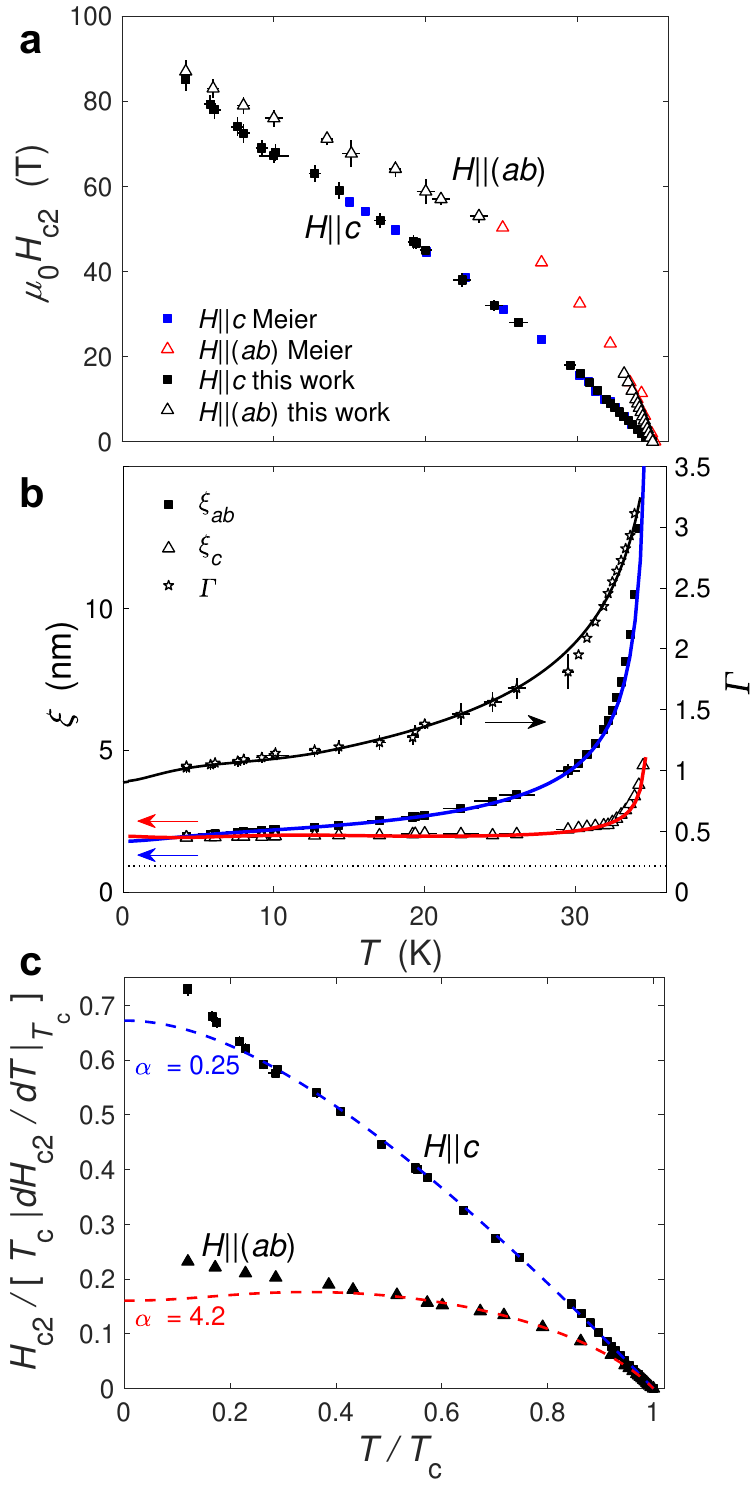}
	\caption{ (a) Upper critical fields as a function of temperature
 in CaKFe$_{4}$As$_{4}$ when $H||c$ (solid squares) and $H||(ab)$ (open triangles),
  together with previous reported data measured up to $60\,$T from Ref.~\onlinecite{Meier2016}.
(b) The temperature dependence of
coherence length extracted close to $T_{\rm c}$, as described in the SM.
The $H_{\rm{c2}}(T=0)$ value were used to find the zero-temperature coherence lengths.
The horizontal dashed line represents the 3D-2D crossover when coherence length $\sim c/\sqrt{2}$.
The  anisotropy $\Gamma=H_{\rm{c2}}^{ab}/H_{\rm{c2}}^{c}$ on the right axis
approaches $\Gamma=1$ at $\sim4.2\,$K. Solid lines are guide to the eye.
(c) Upper critical fields for $H||c$ (black squares) and $H||(ab)$ (black triangles) scaled
by the superconducting transition temperature, $T_{\rm{c}}$,
 and the slope near $T_{\rm{c}}$ from the WHH model against reduced temperature $T/T_{\rm{c}}$.
 The dashed blue and red lines are fits to the
 WHH model for  $H||c$ and $H||(ab)$ using different values of $\alpha $.
 Deviation from the WHH model occurs at low temperatures for both field orientations.}
	\label{Fig2}
\end{figure}

{\bf Upper critical field phase diagram.}
Based on these experimental data, we have constructed the complete upper critical field phase diagram of CaKFe$_{4}$As$_{4}$
down to $4.2\,$K as shown in Fig.~\ref{Fig2}(a) for the two orientations.
Our results are in good agreement with previous studies up to $60\,$T for both offset and onset critical fields \cite{Meier2016}  (see Fig.~\ref{FigSM_TwoBandModels}(a) and (b))
and reveal extremely large upper critical fields, reaching almost $\sim3T_{\rm{c}}$ at the lowest temperatures.
These values are above the Pauli paramagnetic limit,
$\sim1.85T_{\rm c}$, estimated to be $ \sim 65\,$T for a single-band
superconductor and assuming $g$=2 and the weak coupling limit.
The anisotropy of the upper critical field, defined as the ratio of the upper critical
field for different orientations,  $\Gamma=H_{\rm{c2}}^{ab}/H_{\rm{c2}}^{c}$ ,
 drops dramatically with decreasing temperature from $\sim 4$ to $1$,
as shown in Fig.~\ref{Fig2}(b).
Interestingly, the upper critical fields for the two orientations cross at
$T\sim4.2\,$K, leading to isotropic superconducting behaviour in the low temperature limit.
This phenomenon has been found in optimally doped iron-based superconductors,
such as FeSe$_{0.5}$Te$_{0.5}$ \cite{Serafin2010,Braithwaite2010} and
(Ba,K)Fe$_2$As$_2$ \cite{Yuan2009}.
This behaviour reflects the large Pauli paramagnetic effects in iron-based superconductors
and the influence of Fermi surface details on limiting the orbital effects \cite{Yuan2009}.
The Fermi surface of CaKFe$_{4}$As$_{4}$ has significant warping for the outer electron and hole band
that can potentially allow circulating currents out of plane (see Fig. \ref{FigSM:FS}(d) in the Appendix).
Furthermore, the calculated anisotropy
of the penetration depth based on
plasma frequencies (as in Ref.\onlinecite{Hashimoto2010})
is $\Gamma$=$\lambda_{c}/\lambda_{ab} \sim 4.5$, (see Fig.~\ref{FigSM:FS} in the Appendix)
similar to the measured anisotropy close to $T_{\rm c}$ (Fig.~\ref{Fig2}(b)),
suggesting that the Fermi surface details
 play an important role in understanding its superconducting properties.

Having experimentally determined the upper critical fields
for different orientations in magnetic field, we can extract
the associated coherence lengths in different temperature regimes, as detailed in the Appendix.
In the vicinity of $T_c$, the Ginzburg-Landau coherence lengths shown
in Fig.~\ref{Fig2}(b) are $\xi_{ab}^{\rm{GL}}=1.66(4)\,$nm and $\xi_{c}^{\rm{GL}}=0.53(2)\,$nm for CaKFe$_{4}$As$_{4}$,
similar to previously reported values \cite{Meier2016}.
In the low temperature limit,
we find coherence lengths of $\xi_{ab}=1.83(3)\,$nm and $\xi_{c}=1.87(6)\,$nm, as shown in Fig.~\ref{FigSM_CoherenceLengthExtraction} in the Appendix.
This demonstrates the presence of an isotropic
 superconducting state at lowest temperature in CaKFe$_{4}$As$_{4}$.
The coherence lengths approach the 3D to 2D crossover close to $ c/\sqrt{2} \sim0.911\,$nm,
as shown in Fig.~\ref{Fig2}(b) \cite{Suetin2017,Iyo2016}).
The low temperature extracted $\xi$ values  are larger than $\sim0.7\,$nm
reported from STM measurements, which probe locally the vortex lattice
rather than an overall averaged effect \cite{Fente2018}.
Knowing the coherence length, allows us to estimate the depairing current density, $J_{\rm{d}}$,
as described
in the Appendix \cite{Tinkham_Superconductivity}.
Previous magnetisation data
reported a value of $\mu_{0} H_{\rm{c1}}=22(1)\,$mT when $H||c$ at low temperature which,
combined with $H_{\rm{c2}}$ reported here, give $\kappa~99(2)$ and $\lambda=183(6)\,$nm.
Using this value for the penetration depth and $\xi=1.86(3)\,$nm gives
$J_{\rm{d}}=1.61(9)\times10^8\,\rm{A/cm^2}$, which is one of the largest among iron-based superconductors.
This supports the extremely large critical current densities, reported previously for CaKFe$_{4}$As$_{4}$ in Ref.~\cite{Singh2018}.
 The small coherence length of CaKFe$_{4}$As$_{4}$
 is compatible with the presence of the large upper critical field,
 consistent with small Fermi velocities and high $T_{\rm c}$  values of CaKFe$_{4}$As$_{4}$
 ($\xi \sim \hbar v_{\rm F}$/($2 \pi k_{\rm B} T_{\rm c}$) \cite{gurevich2010upper}).
Surface superconductivity can survive in a thin layer of thickness $\sim\xi$ in systems with clean surfaces and it can lead 
to a critical field larger than $H_{\rm{c2}}$.
 In order to establish the importance of those effects, future angular-dependent studies
 would be necessary to identify the role played by surface superconductivity in CaKFe$_4$As$_4$ close to $T_{\rm{c}}$.
Using the extrapolated zero-temperature normal state resistivity $\rho_{0}\sim9.4\,\mathrm{\mu\Omega{}cm}$ in
Fig.~\ref{Fig1}(f) and carrier concentrations from Ref.~\onlinecite{Meier2016},
we can estimate a mean free path of $\ell= 26.6\,$nm.
Since the mean free path due to elastic scattering from impurities is far larger than the coherence length,
$\xi$(0)$ \ll  \ell$, CaKFe$_{4}$As$_{4}$ can be described as being in the clean limit.
Resistivity data in Fig.~\ref{Fig1}(f) show also a sharp superconducting transition indicating a high quality single crystal.
Furthermore, the extrapolated high-field resistivity at low temperatures (see Fig. \ref{FigSM:FS}(d) in the Appendix)
displays a $T^{2}$ dependence indicative of a Fermi-liquid behaviour that extends
 up to $\sim55\,$K.

In order to assess the role of orbital and Pauli paramagnetic effects on the upper critical field of CaKFe$_{4}$As$_{4}$
we first describe the temperature dependence of  the upper critical field using the
three-dimensional Werthamer-Helfand-Hohenberg (WHH) model \cite{Werthamer1966},
with the inclusion of spin paramagnetism effects.
The slope close to $T_{\rm c}$ ($H'_{\rm c2}=-|\textrm{d}H_{c2}/\textrm{d}T |_{T=T_{\rm{c}}} $)
is used to estimate
 the zero-temperature orbital upper critical field  and
$H^{\rm{orb}}_{\rm c2} = 0.73 H'_{\rm c2} $ for the clean limit
($0.69 H'_{\rm c2} $ in the dirty limit) for a single-band weak-coupling superconductor with ellipsoidal
Fermi surface \cite{Helfand1966}.
We find that the orbital pair breaking dominates  the temperature dependence
 for $H||c$ down to 10~K, below which it deviates, as shown in Fig.~\ref{Fig2}(c).	
 However, when the magnetic field is aligned along the conducting $(ab)$
plane, a Pauli pair breaking contribution has to be included which reduces
the orbital-limited critical field by $\mu_{0} H_{\rm{P}} = \mu_{0} H_{\rm{c2}}^{\rm{orb}}/\sqrt{1 + \alpha^2}$, where $\alpha$ is the Maki parameter.
The extracted Maki parameter $\alpha$  is small
 where the orbital effects dominate ($\alpha \sim 0.25$ for $H||c$), but
it becomes significant reaching ${\alpha}=4.2$ for $H||(ab)$ (see Fig.~\ref{Fig2}(c)).
This value is close to that of FeSe$_{0.6}$Te$_{0.6}$ single crystals
where $\alpha\sim5.5$  suggesting that
the upper critical field is dominated by Pauli paramagnetic effects \cite{Khim2010}.
Strong paramagnetic effects are an important
signature of optimally doped iron-based superconductors
\cite{Tarantini2011,Cho2011,Gurevich2011a,Maiorov2014,Fuchs2009}.
For a clean isotropic single-band, the Maki parameter is
given by $\alpha= \pi^2 \Delta$/($4 E_{\rm F}$).
The values of the measured superconducting gap
$\Delta$ varies between $2.4-13\,$meV
whereas the Fermi energies vary significantly for different bands,
being smallest for the inner hole band, $\alpha$ ($\sim3\,$meV)
and the shallow electron band $\delta$ ($\sim10\,$meV) \cite{Khasanov2018,Mou2016}.
The band structure calculation predicts four electron pockets centred at the $M$
point, as shown in Fig.~\ref{FigSM:FS} in the Appendix.
Experimentally, only one electron pocket can be resolved in experiments
due to a significant intrinsic linewidth and the fact that the
bottom of these bands is located very close to Fermi level \cite{Mou2016}.
Thus, the presence of the shallow bands,
together with the small Fermi energies $ E_{\rm F}$
and large superconducting gap $\Delta$ create conditions for
large $\alpha$ and Pauli pair-breaking.

{\bf Upper critical field described by a two-band model.}
In order to describe the complete temperature dependence of $H_{\rm{c2}}(T)$ for CaKFe$_{4}$As$_{4}$, including the upturn below $\sim 10\,$K for both orientations, a two-band model in the clean limit is considered, as detailed in Ref.~\onlinecite{gurevich2010upper} and in the Appendix.
This model accounts for the presence of two different bands, with interband scattering ($\lambda_{11}, \lambda_{22}$) and intraband effects ($\lambda_{12}, \lambda_{21}$) and includes paramagnetic effects and allows for the presence of an FFLO inhomogeneous state at high-fields and low temperatures
 \cite{gurevich2010upper}.
An FFLO state is characterized by a real-space modulation of the superconducting order parameter either in
amplitude or phase such that the system energy is minimized under the constraints of a large Zeeman
energy and superconducting condensation energy \cite{Song2019}.
As the FFLO wave vector $Q$ appears spontaneously below a certain temperature,
$T_{\rm{FFLO}}$, the spinodal instability line in $H_{\rm{c2}}(T)$ at a finite $Q$ acquires the characteristic upturn \cite{gurevich2010upper}.

\begin{figure}[htbp]
	\centering
	\includegraphics[trim={0cm 0cm 0cm 0cm}, width=0.95\linewidth,clip=true]{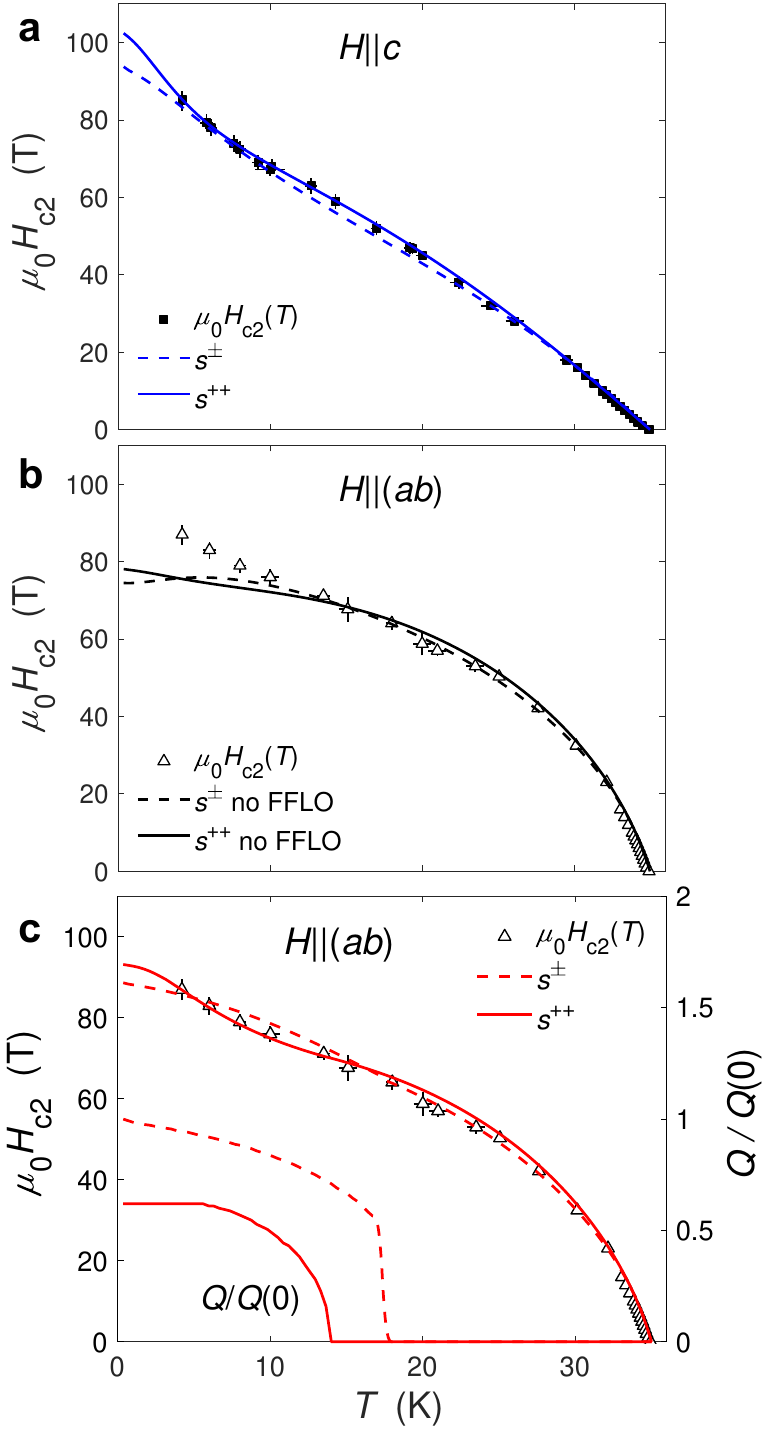}
	\caption{Upper critical fields for (a)
		$H||c$ (solid squares) and (b) and (c)
		$H||(ab)$ (open triangles)
		described by the two-band model.
			The dotted lines correspond to $s^{\pm}$ pairing ($\lambda_{11}=\lambda_{22}=0$, $\lambda_{12}=\lambda_{21}\sim0.5$, $\alpha_{1}=0.95$,
	$\alpha_{2}=0$ for $H||c$ and  $\alpha_{1}=3.1$ and $\alpha_{2}=0.7$ for $H||$($ab$)).
Solid lines represent $s^{++}$ pairing
		($\lambda_{11}=0.81$, $\lambda_{22}=0.29$ and $\lambda_{12}=\lambda_{21}=0.1$, $\alpha_{1}=0.5$ for $H||c$
		and  $\alpha_{1}=3.2$, $\alpha_{2}=0$ for $H||$($ab$)).
		The values of  $\eta$ vary between $\sim 0.02$ in (a)  and $\sim 0.04$ in (b).
		(c) The temperature dependence of the upper critical field
		including the emergence of the FFLO state for the two different pairing symmetries
		with the same parameters as in (b).
		The lower dashed and red line
		 represents the scaled FFLO $Q/Q(0)$ modulation for $H||$($ab$)
		 for the respective pairing (right axis).}
	\label{Fig3}
\end{figure}

In most iron-based superconductors,
the pairing is expected to be mediated by spin-fluctuations
leading to a sign changing  $s_{\pm}$ order parameter.
In this case, upper critical fields
 are generally described by dominant interband coupling parameters with $\lambda_{11}=\lambda_{22}=0$ and $\lambda_{12}=\lambda_{21}\sim0.5$
  ($\lambda_{11}\lambda_{22} \ll \lambda_{12}\lambda_{21}$) \cite{Hirschfeld2011, Cho2011,Tarantini2011,Meier2016,Kano2009,Bristow2019_FeSeSHc2}.
Furthermore, in these multiband systems orbital fluctuations can dominate the pairing interactions in certain conditions
favouring a gap with equal sign on each pocket, resulting in the $s^{++}$ pairing  \cite{Kontani2010}
 ($\lambda_{11}\lambda_{22} \gg \lambda_{12}\lambda_{21}$).

To model the upper critical fields for both field orientations, we have considered different
input parameters from previous experiments on CaKFe$_{4}$As$_{4}$.
For example, the ratio between the extreme velocities of the Fermi surfaces is taken from ARPES experiments \cite{Khasanov2018} and
it gives a starting value of $\eta_{c}\sim0.02$,
which is the squared velocity ratio between the two bands.
Upper critical field simulations have a strong sensitivity to $\eta$ values,
as shown in Fig.~\ref{FigSM_TwoBand_Simulations} in the Appendix
and it can change for different field orientations due to differences
in the in-plane velocities on the Fermi surface (see Figs.~\ref{Fig1}(e) and \ref{FigSM:FS} in the Appendix).
We assume that $s$=1 (defined by Eq.~\ref{TwoBandFFLO_Params_equ} in the Appendix) for both field orientations
and the starting values of the
Maki parameters are those
from the single band model of $H_{\rm{c2}}(T)$
in Fig.~\ref{Fig2}(c).

Firstly, the $H_{\rm{c2}}(T)$ data are simulated for CaKFe$_{4}$As$_{4}$
using parameters corresponding to the $s^{\pm}$ case (dashed lines Figs.~\ref{Fig3}(a) and \ref{Fig3}(b)),
which provide a reasonable representation of the observed behaviour for both orientations.
However, we find that $H_{\rm{c2}}(T)$ is best described
using $\lambda_{11}=0.81$, $\lambda_{22}=0.29$ and $\lambda_{12}=\lambda_{21}=0.1$
(solid lines in  Figs.~\ref{Fig3}(a) and \ref{Fig3}(b)),
which is consistent with $s^{++}$ pairing.
For the later model, the constrained  Fermi velocities are
 $v_{1}\sim380(20)\,\mathrm{meV\AA}$ and $v_{2}\sim54(4)\,\mathrm{meV\AA}$ when $H||c$, and $v_{1}\sim195(20)\,\mathrm{meV\AA}$ and $v_{2}\sim41(4)\,\mathrm{meV\AA}$ when $H||(ab)$ (see Table~\ref{CaKFe4As4_TwoBand_Table}).
The values of $v_{\rm{2}}$ are similar in both orientations, indicative of a strongly warped sheet (see Fig.~\ref{FigSM:FS} in the Appendix).
On the other hand, $v_{1}$ is larger for $H||c$
and is close to  $\sim360\,\mathrm{meV\AA}$, reported for
 the $\alpha$ hole band from ARPES measurements \cite{Khasanov2018}.
The band coupling constants, $\lambda_{\rm i,j}$ that describe
our data suggest that  the interband scattering due to spin-fluctuations
is dominated by the intraband effects in  CaKFe$_{4}$As$_{4}$,
which can occur in the presence of
strong orbital fluctuations \cite{Kontani2010}.
For  $s^{++}$ pairing to overcome  $s^{\pm}$ pairing
it normally requires the presence of strong disorder which
leads to similar superconducting gaps on
different  Fermi  surfaces \cite{Efremov2011}.
Furthermore, the band coupling constants
for the  $s^{++}$ case resemble those 
used to describe a two-band superconductor in the dirty limit \cite{Golubov2002}.
However, CaKFe$_{4}$As$_{4}$ is clean, suggesting that disorder effects are negligible.
Thus, based on the band coupling parameters, we conclude that orbital fluctuations may dominate over spin fluctuations.

When the magnetic field lies along the conducting planes,
the low temperature upturn of  $H_{\rm{c2}}(T)$ in  Fig.~\ref{Fig3}(b)
cannot be fully captured by only considering the two-band model with
large Pauli paramagnetic effects, $\alpha \sim 3.2$.
The formation of the FFLO state in a system with a cylindrical Fermi surface
requires a large Zeeman energy and a critical Maki's parameter of $\alpha_{\rm c}=4.76$ \cite{Song2019}, compared to  $\alpha_{\rm c}=1.8$ \cite{Gruenberg1966}
for a three-dimensional Fermi surface.
CaKFe$_{4}$As$_{4}$ has a complex Fermi surface
 with two-dimensional cylindrical and highly warped sheets
and together with a large value of $\alpha$
creates the conditions for the emergence of an FFLO state \cite{gurevich2010upper}.
A FFLO state  can be realized in very clean materials with weak scattering of quasiparticles
and it generally manifests as change in slope in the upper critical field at low temperatures  \cite{gurevich2010upper}.
We find indeed that in order to describe the upper critical field data of CaKFe$_{4}$As$_{4}$  over
the entire temperature range, a FFLO state could be stabilized below $T_{\rm{FFLO}}\sim14\,$K, as shown
in Fig.~\ref{Fig3}(c). Possible contenders to support such an effect are
 the shallow electron Fermi surface pocket, $\delta$, in the
zone corner which is very close to the Fermi level ($\sim10\,$meV)
as well  as the inner hole band $\alpha$ ($\sim3\,$meV),
as detected by ARPES measurements  \cite{Mou2016}.

The temperature dependence
of the upper critical field data in CaKFe$_{4}$As$_{4}$
implies that the intraband scattering
is likely to dominate the interband scattering,
the latter of which being promoted by spin fluctuations.
In the presence of spin-orbit coupling
the orbital fluctuations can lead to $s^{++}$ pairing,
as suggested for LiFeAs \cite{Saito2015}.
Usually $s^{++}$ pairing results in far
lower critical fields than $s^{\pm}$ pairing \cite{gurevich2010upper},
so our results for  CaKFe$_4$As$_4$
are somehow unusual as $H_{\rm{c2}}$  is almost $3T_{\rm{c}}$ at the lowest temperature.
The presence of several scattering channels in a multi-band system like
CaKFe$_4$As$_4$ can increase the upper critical field to a much
greater extent than in single gap superconductors, caused
by the relative weight of different scattering channels.
CaKFe$_4$As$_4$ theoretically has up to six large cylindrical hole bands,
(with the equivalent of 50\% hole doping in a 122 iron-based superconductor)
providing a large density of states (see Fig.~\ref{Fig1}(e), Fig.~\ref{FigSM:FS} in the Appendix and Ref.~\onlinecite{Khasanov2018}) and it can also promote intraband scattering
driven by orbital or electron-phonon couplings.
On the other hand, the shallow bands in CaKFe$_4$As$_4$
are likely to be involved in the stabilization of the FFLO state.
Thus, the temperature dependence of the upper critical field of CaKFe$_4$As$_4$
can reflect the behaviour of two dominant superconducting gaps
that reside on different sheets of the cylindrical Fermi surface,
possibly one on a large hole band and another one on a shallow
small band.

In summary, we have experimentally mapped the upper critical fields of CaKFe$_{4}$As$_{4}$ up to $90\,$T and down to $4.2\,$K, providing
a complete phase diagram for this stoichiometric superconductor. The anisotropy decreases
dramatically with temperature, the system becoming essentially isotropic near $4\,$K.
Upper critical fields are extremely large in this system, reaching close to $\sim3T_{\rm{c}}$ at the lowest temperature,
well above the expectation based on conventional single-band superconductivity.
Instead, the temperature dependence of the upper critical field can be described using
by a two-band model in the clean limit.
The band coupling constants suggest a dominant orbital pairing over spin fluctuations pairing in CaKFe$_{4}$As$_{4}$.
Furthermore,
for magnetic fields aligned in the conducting plane
and due to large Pauli paramagnetic effects,
we find that the temperature dependence of the upper critical field is
 consistent with the emergence of an FFLO state at low temperatures.

{\bf Acknowledgements}
We thank  Alex Gurevich for helpful  discussions
related to the two-band modelling of the upper critical field
and J\'{e}r\^{o}me B\'{e}ard, Marc Nardone, Abdelaziz Zitouni
for the technical support during the pulsed field experiments.
This work was mainly supported by Oxford Centre for Applied Superconductivity.
P.R. acknowledges the support of the Oxford Quantum Materials Platform Grant (EP/M020517/1).
Part of this work was supported
HFML-RU/FOM and LNCMI-CNRS, members of the European Magnetic Field
Laboratory (EMFL) and by EPSRC (UK) via its membership to the EMFL
(grant no. EP/N01085X/1).
Part of this work at the LNCMI was supported by Programme
Investissements d'Avenir under the program ANR-11-IDEX-0002-02, reference ANR-10-LABX-0037-NEXT.
Work at Ames Laboratory was supported by the U.S. Department of Energy, Office of Basic Energy Science, Division of Materials Sciences and Engineering. Ames Laboratory is operated for the U.S. Department of Energy by Iowa State University under Contract No. DE-AC02-07CH11358. W.M. was supported by the Gordon and Betty Moore Foundation’s EPiQS Initiative through Grant GBMF4411.
We also acknowledge financial support of the John Fell
Fund of the Oxford University.
AIC acknowledges an EPSRC Career Acceleration Fellowship (EP/I004475/1)
and Oxford Centre for Applied Superconductivity.

\bibliography{Ca1144Hc2_bib_oct19}

\section{Appendix}

In this section we present additional data to support the findings presented in the main paper.

\begin{figure*}[h!]
	\centering
	\includegraphics[width=0.85\linewidth,clip=true]{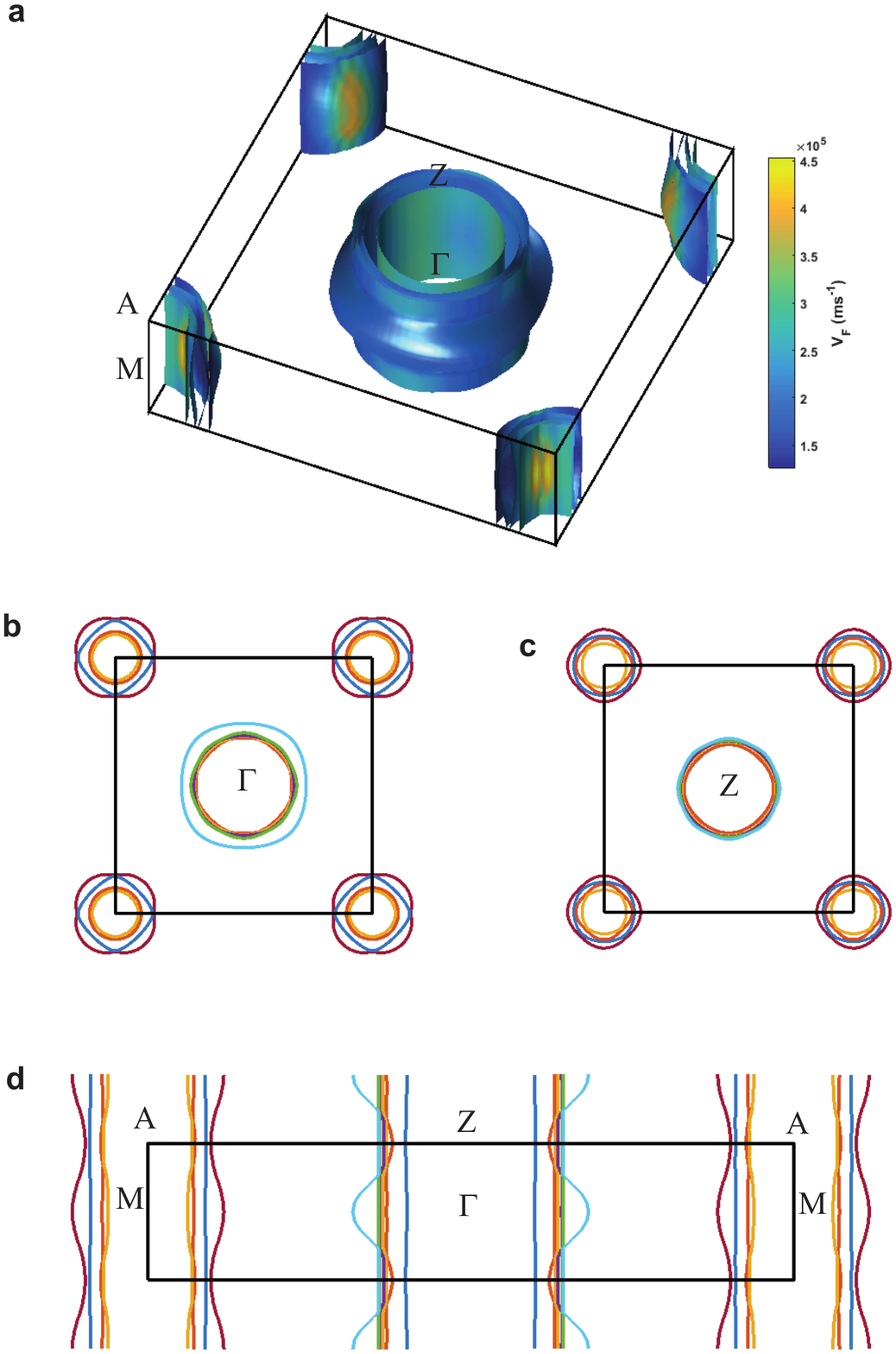}
	\caption{\textbf{Fermi surface of CaKFe$_{4}$As$_{4}$.}
		a) Fermi surface with velocity colour. Slices at the b) $\Gamma$ and b) $Z$ point indicating the size of the different sheets.
		Solid lines indicate the Brillouin zone.
		c) Fermi surface slice along the [110] diagonal of the Brillouin zone indicated by the solid lines.
		The calculations were performed using Wien2k and GGA approximation
		and the experimental lattice parameters described by the  $P4/m m m$ symmetry group
		were $a$=$b$=3.86590~\AA, $c$=12.88400~\AA \cite{Iyo2016}.
		The calculated value of anisotropy, $\Gamma$=$\lambda_{c}/\lambda_{ab}$=313.68/70=4.5
		using a similar approached used in Ref.\onlinecite{Hashimoto2010}.
		These calculations are in agreement with previous results reported in Ref. \onlinecite{Suetin2017}.
	}
	\label{FigSM:FS}
\end{figure*}

\begin{figure*}[htbp]
	\centering
	\includegraphics[width=0.75\linewidth,clip]{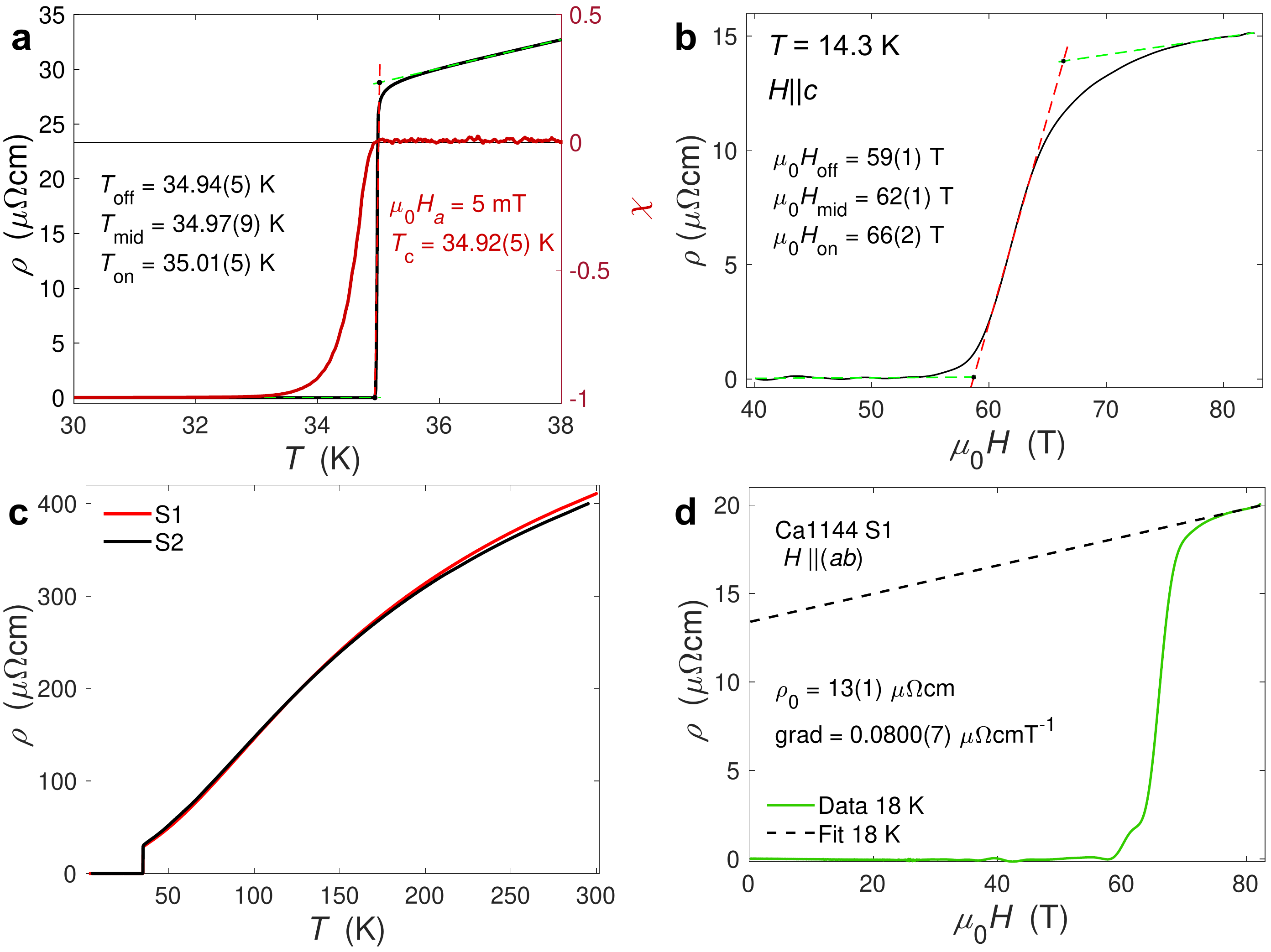}
	\caption{\textbf{Transport data of CaKFe$_{4}$As$_{4}$ as a function of temperature and magnetic field.}
	(a) Estimating the critical temperature, $T_{\rm{c}}$, from resistivity against temperature in zero magnetic field (left-axis) and from magnetic susceptibility in a small applied field of $5\,$mT (right-axis).
	The midpoint of the transition, $T_{\rm{mid}}$, is identified as the peak in the derivative $d\rho/dT$.
	Linear fits are performed above and below the midpoint transition and the intercepts are defined as $T_{\rm{off}}$ and $T_{\rm{on}}$, as labelled in the figure.
	$T_{\rm{c}}$ from susceptibility is defined as the point at which diamagnetism occurs and deviates from the high-temperature background and it is closest to the offset temperature in transport.  The critical temperatures used to build the phase diagrams in magnetic field were extracted using $T_{\rm{off}}$ from the transport data.
	(b) The estimation of the upper critical field, $H_{\rm{off}}$, from resistivity in pulsed magnetic field data measured at constant temperature using the same method as in (a).
	(c) Resistivity against temperature for different samples of CaKFe$_{4}$As$_{4}$ between $300\,$K and $2\,$K.
	The superconducting transition is very sharp, with $T_{\rm{c,0}}=35.0\,$K and width $\Delta T_{\rm{c}}=0.15\,$K,
	values typical for all measured single crystals indicating their high quality.
	(d) Resistivity data as a function of magnetic field for the sample S1 when $H||(ab)$ at $18\,$K.
	The dashed black line shows the linear extrapolation used to obtain the zero-field resistivity at 18~K
	and the intercept, $\rho_{H\rightarrow0}$($T$), and gradient are shown.
	Using this method, the extracted values of zero-field resistivitiy measured at constant temperatures are shown in the main paper in Fig.~\ref{Fig1}(f).	}
	\label{FigSM_rhoT}
\end{figure*}

\newpage
\clearpage
\vspace{0.5cm}

{\bf A. Experimental details.}

\begin{figure}[htbp]
	\centering
	\includegraphics[width=1\linewidth,clip]{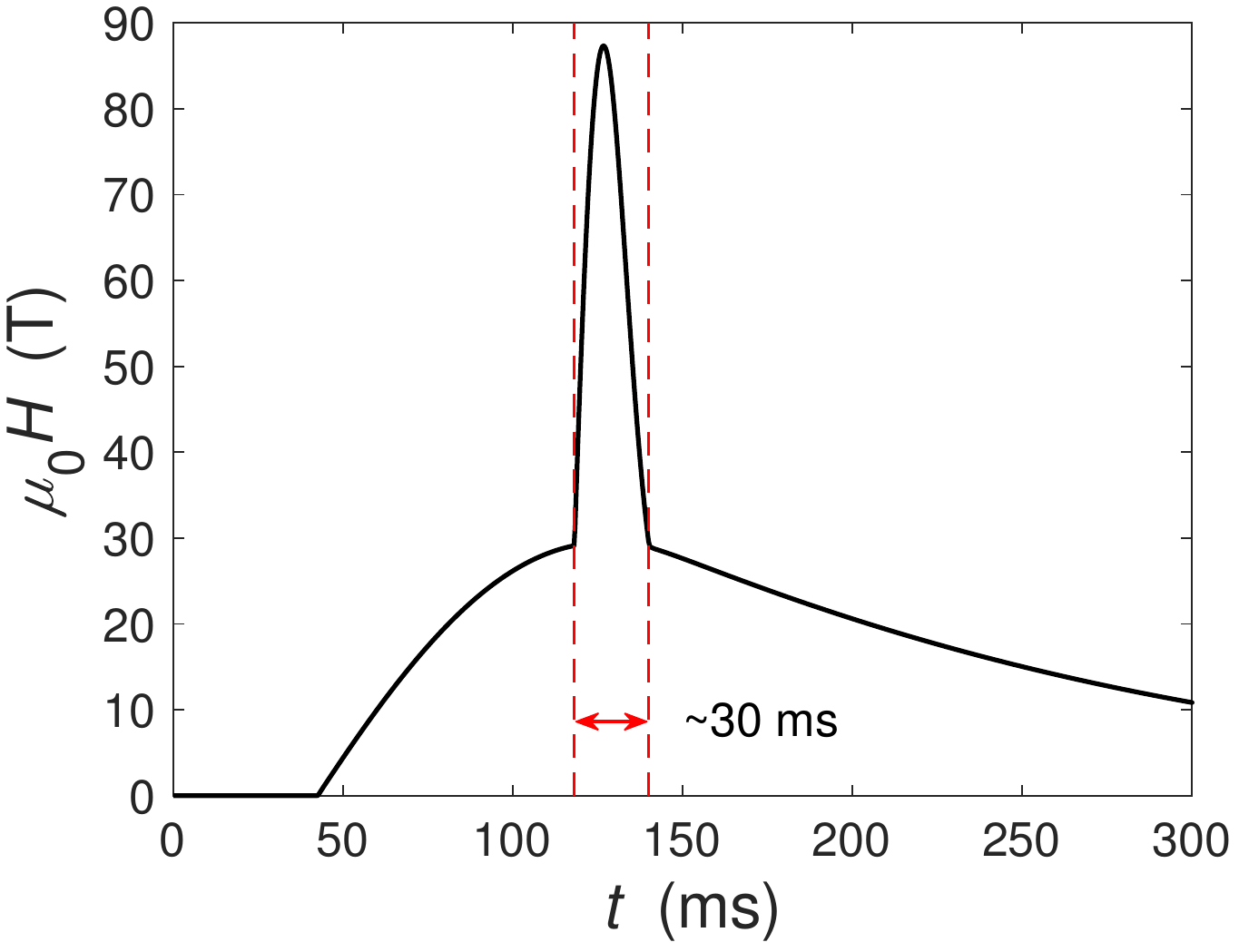}
	\caption{\textbf{Generation of the high-magnetic field pulse.}
		Applied magnetic field as a function of time from the $90\,$T magnet in Toulouse. The peak of this pulse was $\sim88\,$T and is generated by overlaying pulsed current though two different coils. The pulse on the inner coil which generates the extremely high fields lasts $\sim30\,$ms.
	}
	\label{FigSM_PulseExample}
\end{figure}

Several high quality single crystals were used in this study
with large residual resistivity ratios $RRR=\rho(300\rm{K})/\rho(36\rm{K})\sim14.5$,
small $\rho_{\rm{0}}$ values of $\sim9.2\,\mathrm{\mu\Omega{}cm}$
and sharp superconducting transitions with $\Delta T_{\rm{c}} \sim 0.1\,$K, as shown in
Figs.~\ref{FigSM_rhoT}(c) and Fig.1(f).
Temperature sweeps were performed at constant magnetic field
for different orientations, $H||c$ and $H||(ab)$ to build the high-temperature part of the $H_{\rm{c2}}(T)$ phase diagram.
Magnetic fields up to $90\,$T were produced by using current pulses through two different solenoid coils.
An example of the magnetic field produced as a function of time in shown in Fig.~\ref{FigSM_PulseExample}.

\vspace{0.5cm}

{\bf B. Determining superconducting coherence lengths.}

The superconducting coherence lengths can be found in both field orientations near $T_{\rm{c}}$ using the Ginzburg-Landau (GL) approach from the equations:
\begin{equation}
\xi_{ab}^{\rm{GL}}=\sqrt{\frac{\Phi_{\rm{0}}}{2\pi\mu_{\rm{0}}\lvert\frac{ dH_{\rm{c2}}^{c}}{dT}\rvert _{T_{\rm{c}}}T_{\rm{c}}}},
\;\;
\xi_{c}^{\rm{GL}}=\frac{\Phi_{\rm{0}}}{2\pi\mu_{\rm{0}}\lvert\frac{ dH_{\rm{c2}}^{ab}}{dT}\rvert _{T_{\rm{c}}}T_{\rm{c}}\xi_{ab}^{\rm{GL}}},
\label{GL_xi_Equations}
\end{equation}
where $\Phi_{\rm{0}}$ is the magnetic flux quanta, $\mu_{\rm{0}}\lvert dH_{\rm{c2}}^{c}/dT\rvert _{T_{\rm{c}}}$ is the slope near $T_{\rm{c}}$ when $H||c$, $\mu_{\rm{0}}\lvert dH_{\rm{c2}}^{ab}/dT\rvert _{T_{\rm{c}}}$ is the slope near $T_{\rm{c}}$ when $H||(ab)$ and $\xi_{ab}^{\rm{GL}}$ and $\xi_{c}^{\rm{GL}}$ are the respective GL coherence lengths.

At the lowest temperature,
the coherence lengths are extracted using
$\mu_{\rm{0}}H_{\rm{c2}}(T\rightarrow0)$ and the following equations:
\begin{equation}
\xi_{ab}=\sqrt{\frac{\Phi_{\rm{0}}}{2\pi\mu_{\rm{0}}H_{\rm{c2}}^{c}(T\rightarrow0)}},
\;\;
\xi_{c}=\frac{\Phi_{\rm{0}}}{2\pi\mu_{\rm{0}}H_{\rm{c2}}^{ab}(T\rightarrow0)\xi_{ab}},
\label{xi_lowT_Equations}
\end{equation}
where $\xi_{ab}$ and $\xi_{c}$ are the coherence lengths in the ($ab$)-plane and along the $c$-axis, and $\mu_{\rm{0}}H_{\rm{c2}}^{c}(T\rightarrow0)$ and $\mu_{\rm{0}}H_{\rm{c2}}^{ab}(T\rightarrow0)$ are the critical fields at the lowest temperature when $H||c$ and $H||(ab)$ respectively.
Using equations~\ref{xi_lowT_Equations} at different constant temperatures, one can estimate
$\xi_{i}\rightarrow\xi_{i}(T)$ and $H_{\rm{c2}}\rightarrow H_{\rm{c2}}(T)$,
and the anisotropy ratio, $\Gamma$, as shown in Fig.2(b) in the main text.

\begin{figure}[htbp]
	\centering
	\includegraphics[width=1\linewidth,clip=true]{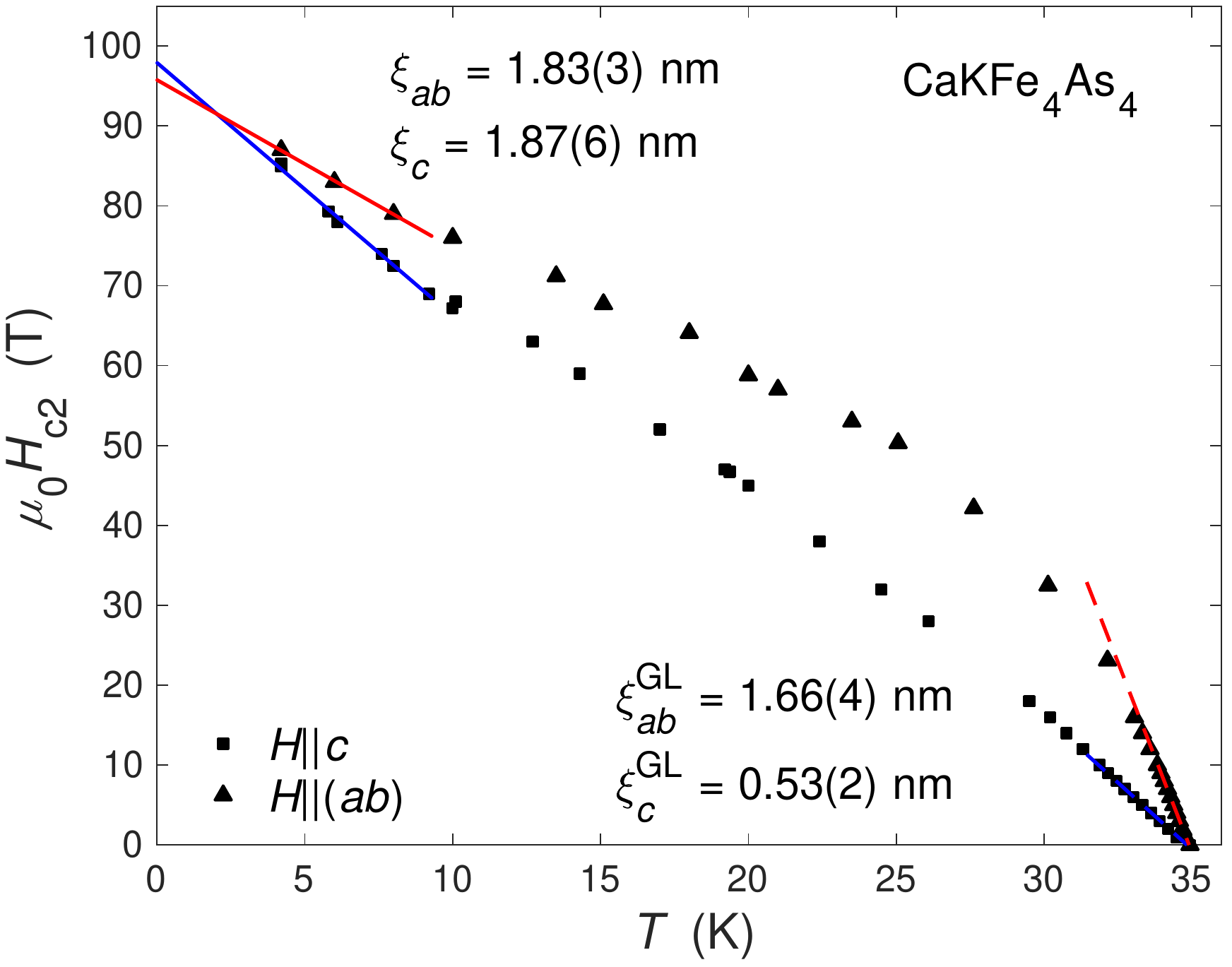}
	\caption{\textbf{Coherence length extraction.}
		Upper critical fields in CaKFe$_{4}$As$_{4}$ as a function of temperature.
		The dashed blue and red lines are linear fits near $T_{\rm{c}}$ when $H||c$ and $H||(ab)$ respectively. The gradients of these fits (shown in the figure)
		were used to calculate the coherence lengths using the Ginzburg-Landau theory (equation~\ref{GL_xi_Equations}).
		The solid blue and red lines are linear fits to the low temperature region when $H||c$ and $H||(ab)$ respectively.
		The zero-temperature extrapolation of $H_{\rm{c2}}$($T\rightarrow0$)
		for the two orientations were used to find the zero-temperature coherence lengths, as shown in the figure.}
	\label{FigSM_CoherenceLengthExtraction}
\end{figure}

\vspace{0.5cm}

{\bf{C. Depairing current density}}

Using the coherence length, one can also determine the critical depairing current density of a superconductor, $J_{\rm{d}}$.
As a function of temperature and magnetic field, this is be expressed as \cite{Tinkham_Superconductivity}:
\begin{equation}
J_{\rm{d}}(T,H) = \frac{2}{3\mu_{\rm{0}}\lambda^{2}(T)}
\sqrt{\frac{\mu_{0}(H_{\rm{c2}}^{2}(T)-H^{2})^{1/2}\Phi_{0}}{6\pi}},
\label{Jd_equ_1}
\end{equation}
where $\lambda$ is the superconducting penetration depth and $\Phi_{0}$ is the magnetic flux quantum.
At zero temperature and in no magnetic field, this gives
\begin{equation}
J_{\rm{d}}(0,0) = \frac{2}{3\mu_{0}\lambda^{2}(0)}
\sqrt{\frac{\mu_{0}H_{\rm{c2}}(0)\Phi_{0}}{6\pi}}.
\label{Jd_equ_2}
\end{equation}
Substituting $\mu_{0}H_{\rm{c2}}=\Phi_{0}/2\pi\xi^2$ this equation can be rearranged to give
\begin{equation}
J_{\rm{d}}(0,0) = \frac{\Phi_{0}}{3\sqrt{3}\pi\mu_{0}}\frac{1}{\lambda^{2}(0)\xi(0)}.
\label{Jd_equ_3}
\end{equation}

Previous magnetisation reported $\mu_{0} H_{\rm{c1}}=22(1)\,$mT  \cite{Singh2018}
when $H||c$ at low temperature which, combined with $H_{\rm{c2}}$ values reported here,
give $\kappa \sim 99(2)$ and $\lambda=183(6)\,$nm.
Using  $\xi=1.86\,$nm,
we find that the
depairing current density is $J_{\rm{d}}\sim1.62\times10^{8}\,\mathrm{A/cm^{2}}$,
which is among the largest of any known iron-based superconductor.

\vspace{0.5cm}
{\bf D. The upper critical fields using a two-band model.}

To describe the entire temperature dependence of $H_{\rm{c2}}(T)$ for CaKFe$_{4}$As$_{4}$,
including the upturn below $\sim10\,$K for both orientations,
we use a two-band model in the clean limit.
This models includes paramagnetic effects and
allows the emergence of a FFLO state, as detailed in Ref.~\onlinecite{gurevich2010upper}.
The values of the upper critical field for different temperatures are found from estimating
the following expressions:
\begin{multline}
a_{1}[\ln t + U_{\rm{1}}]+a_{2}[\ln t + U_{\rm{2}}] \\
+ [\ln t + U_{\rm{1}}][\ln t + U_{\rm{2}}] = 0, \;\;\;\;
\label{TwoBandFFLO_equ}
\end{multline}
where $a_{\rm{1}}=(\lambda_{\rm{0}}+\lambda_{\rm{-}})/2w$,
$a_{\rm{2}}=(\lambda_{\rm{0}}-\lambda_{\rm{-}})/2w$,
$\lambda_{\pm}=\lambda_{11}\pm\lambda_{22}$,
$w=\lambda_{11}\lambda_{22}-\lambda_{12}\lambda_{21}$ and
$\lambda_{\rm{0}}=(\lambda_{-}^{2}+4\lambda_{12}\lambda{21})^{\rm{1/2}}$.
Here $\lambda_{ij}$ represents a coupling constant between bands $i$ and $j$, and as the cross-terms $\lambda_{12}$ and $\lambda_{21}$ only appear together we set $\lambda_{21}=\lambda_{12}$ for simplicity.
The values of the coupling constants allows us to establish the dominant pairing.
For $s^{++}$ pairing the parameter $w>0$
($w \sim 0.248$ for the case presented in Fig.~\ref{Fig3}(c) and  Fig.~\ref{FigSM_TwoBandModels})(c))
whereas for $s^{\pm}$ pairing $w<0$
(Fig.~\ref{Fig3}(a) and (b),
and Figs.~\ref{FigSM_TwoBandModels}(a) and \ref{FigSM_TwoBandModels}(b)).

$U_1$ and $U_2$ are defined as
%$U=U(t,b,\alpha,q,\eta,s)$ where
\begin{multline}
U_1 = 2e^{q^{2}}\rm{Re}\it \sum_{n=0}^{\infty}\int_{q}^{\infty}du{}e^{-u^{2}} \\
\Bigg( \frac{u}{n+1/2} - \frac{t}{\sqrt{b}}\rm{tan^{-1}}\Bigg[ \it \frac{u\sqrt{b}}{t(n+\rm{1/2}) + \it i\alpha{}b} \Bigg] \Bigg)
\label{TwoBandFFLO_U1_equ}
\end{multline}
\begin{multline}
U_2 = 2e^{q^{2}s}\rm{Re}\it \sum_{n=0}^{\infty}\int_{q\sqrt{s}}^{\infty}du {}e^{-u^{2}} \\
\Bigg( \frac{u}{n+1/2} - \frac{t}{\sqrt{b\eta}}\rm{tan^{-1}}\Bigg[ \it \frac{u\sqrt{b\eta}}{t(n+\rm{1/2}) + \it i\alpha{}b} \Bigg] \Bigg)
\label{TwoBandFFLO_U2_equ}
\end{multline}

\begin{figure*}[ht]
	\centering
	\includegraphics[width=1.0\linewidth,clip=true]{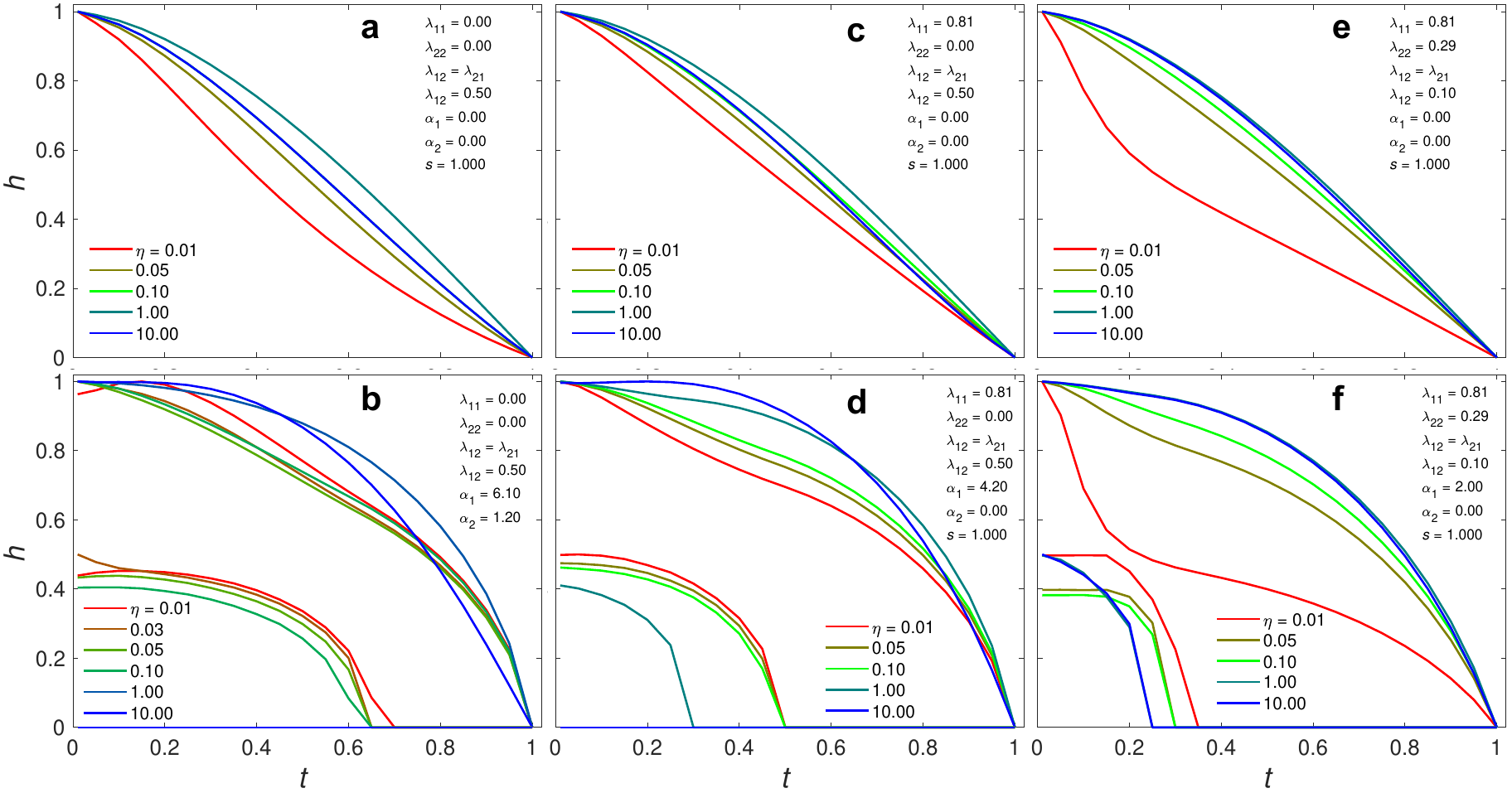}
	\caption{\textbf{Two-band upper critical field simulations for different band coupling parameters and values of $\eta$.}
		Simulations of the reduced upper critical field $h$ using the two-band model based on equations~\ref{TwoBandFFLO_equ}-\ref{TwoBandFFLO_Params_equ}.
		The parameters listed in each panel correspond to $s^{\pm}$ pairing in (a) and (b) with $\lambda_{11}=\lambda_{22}=0$, $\lambda_{12}=\lambda_{21}=0.5$,
		$s^{\pm}$ pairing in (c) and (d) with $\lambda_{11}=\lambda_{22}=0.81$, $\lambda_{12}=\lambda_{21}=0.5$,
		and $s^{++}$ pairing in (e) and (f) with $\lambda_{11}=0.81$, $\lambda_{22}=0.29$ and $\lambda_{12}=\lambda_{21}=0.1$, $\alpha_{1}=0.95$ for $H||c$.
		For panels (b), (d) and (f) the lower curves (left corners) represent the magnitude of the FFLO modulation, $Q/(2Q(0))$.}
	\label{FigSM_TwoBand_Simulations}
\end{figure*}

\begin{figure*}[htbp]
	\centering
	\includegraphics[width=0.75\linewidth,clip=true]{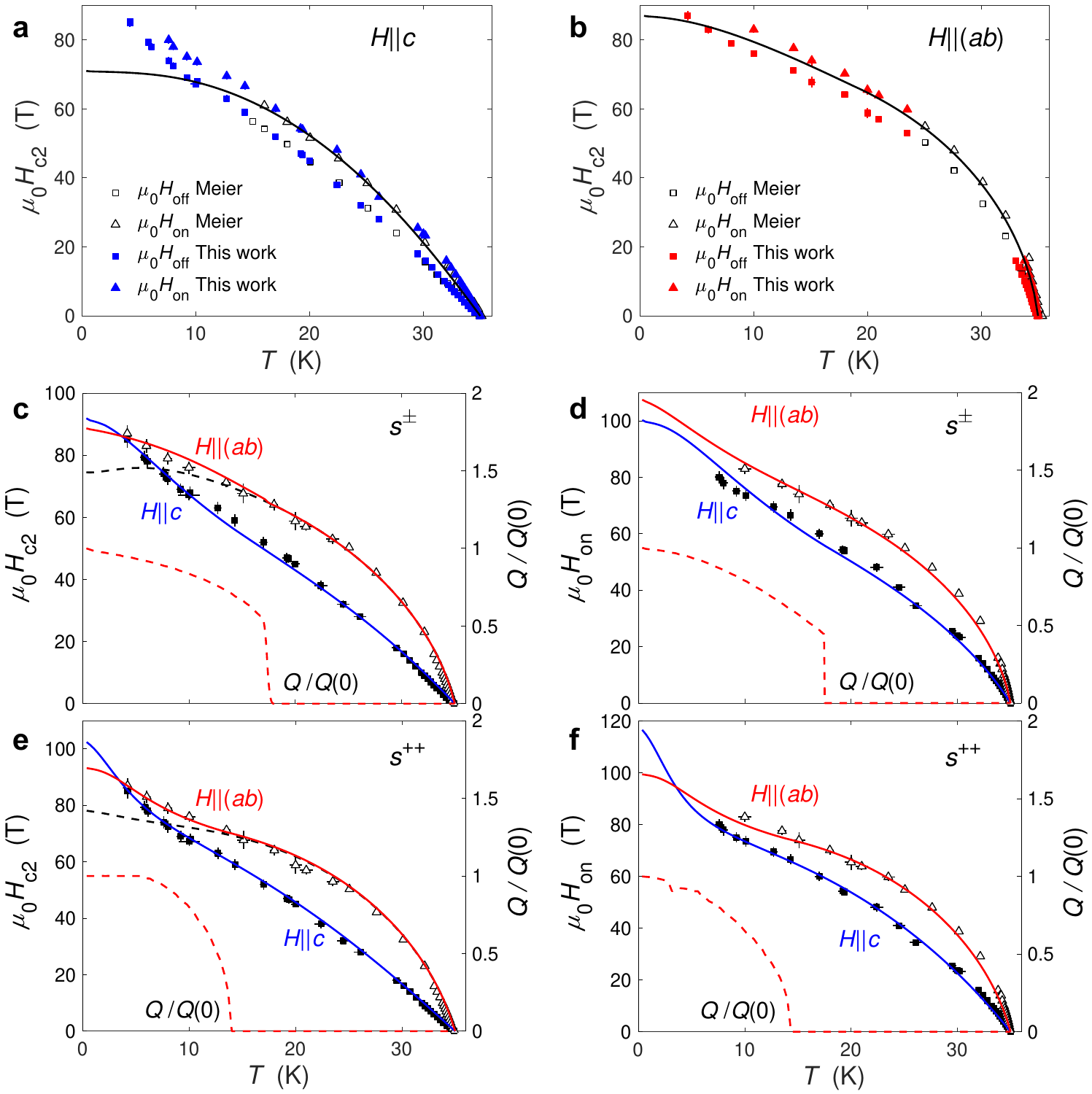}
	\caption{\textbf{Two-band upper critical field models for CaKFe$_{4}$As$_{4}$ using different band coupling parameters.}
		(a) and (b)	$\lambda_{11}=\lambda_{22}=0$, $\lambda_{12}\lambda_{21}=0.25$, $\eta=0.2$, $\alpha=0.5$
		and $\epsilon=1/6$ used previously to describe the onset upper critical field of CaKFe$_4$As$_4$ up to $60\,$T \cite{Meier2016} when $H||c$ and $H||(ab)$ respectively.
		(c)  $s^{\pm}$ pairing with $\lambda_{11}=\lambda_{22}=0$, $\lambda_{12}=\lambda_{21}=0.5$.
		Here $\alpha_{1}=0.95$, $\alpha_{2}=0$ and $\eta\sim0.006$ for $H||c$,
		and $\alpha_{1}=3.1$, $\alpha_{2}=0.7$ and $\eta\sim0.04$ for $H||$($ab$).
		(d) $s^{\pm}$ pairing for the onset critical field using the same parameters as in (c), but with $\eta\sim0.008$ and $\alpha_{1}=1.5$ when $H||c$, and $\alpha_{1}=4.2$ when $H||(ab)$.
		(e) $s^{++}$ pairing with $\lambda_{11}=0.81$, $\lambda_{22}=0.29$ and $\lambda_{12}=\lambda_{21}=0.1$. Here $\alpha_{1}=0.95$, $\alpha_{2}=0$ and $\eta\sim0.02$ for $H||c$,
		and $\alpha_{1}=3.1$, $\alpha_{2}=0.0$ and $\eta\sim0.04$ for $H||(ab)$.
		(f) $s^{++}$ pairing for the onset critical field using the same parameters as in (e), but with $\alpha_{1}=1.0$ when $H||c$, and $\alpha_{1}=3.5$ when $H||(ab)$.
		In (c)-(f) blue lines show the two-band model (with no FFLO state present) when $H||c$, dashed black lines show the two-band model with no FFLO state when $H||(ab)$, and the red lines show the two-band model with an FFLO state when $H||(ab)$.
		The temperature dependence of the FFLO $Q$ vector is shown by the dashed red lines (the right-axis in  (c)-(f)) for $H||(ab)$, and it is scaled to the zero-temperature value $Q(0)$.
		We note that the FFLO state emerges when $H||(ab)$ regardless of the pairing limit or which critical field criteria is chosen. This reinforces the robustness of this result.
	}
	\label{FigSM_TwoBandModels}
\end{figure*}

The variables used for this two-band model, $b$, $\alpha$, $q$, $\eta$ and $s$, are defined as
\begin{multline}
\;\;\;\;\;\;\;\;\;\;\;\;\;\;
b = \frac{\hbar^2 v_{1}^{2}H_{\rm c2}}{8\pi\phi_{\rm{0}}{k_{\rm{B}}}^2\it{T}_{\rm{c}}^{2}}, \;\;\;\;
\alpha = \frac{4\mu\phi_{\rm{0}} {k_{\rm{B}}}\it{T}_{\rm{c}}}{\hbar{}^2 v_{1}^{2}},
\\
q^{2} = \frac{Q^2 \phi_{\rm{0}}\epsilon_{1}}{2\pi H_{\rm c2}}, \;\;\;\;
\eta = \frac{v_{2}^{2}}{v_{1}^{2}}, \;\;\;\;
s = \frac{\epsilon_{2}}{\epsilon_{1}}.
\;\;\;\;\;\;\;\;\;\;\;\;
\label{TwoBandFFLO_Params_equ}
\end{multline}
$v_{i}$ is the in-plane Fermi velocity of band $i$, 
$\epsilon_{i}$ is the mass anisotropy ratio,
$\epsilon_{i}= m^{\perp}_{i}/m^{\parallel}_{i}$ is related to the ratio between the gradients near $T_{\rm{c}}$ in different field orientations,
$s$ is band mass anisotropy between the two bands and
$ \gamma= \epsilon^{-1/2}$.
Various simulations of upper critical fields using this two-band model are shown in Fig.~\ref{FigSM_TwoBand_Simulations}.
The slope of $H_{\rm c2}$ near $T_{\rm{c}}$ of CaKFe$_{4}$As$_{4}$,
is $-3.3$ for $H||c$ and $-10.8$ for $H||$($ab$), giving an $\epsilon\sim1/10$.
$Q$ is the magnitude of the FFLO modulation
and it is found for a given temperature when $H_{\rm{c2}}$ is maximal ($dH_{\rm{c2}}/dQ=0$).
The simulations of $H_{\rm c2}$ for $H||(ab)$  assumed that the bands have the same anisotropy
parameter $\epsilon_{1} = \epsilon_{2} = \epsilon$.
Other parameters are estimated based on the fitted values using a single-band model
and the velocities are minimised locally to find an optimum solution.
This approach was used for the two different $s^{\pm}$ and $s^{++}$ pairing,
which are described in the main body.

A rescaling of the upper critical field has been performed previously mapping $H_{\rm{c2}}^{c}\rightarrow H_{\rm{c2}}^{ab}$ using
$q_{ab} \rightarrow q_{c} \epsilon^{-3/4}$,
$\alpha_{ab} \rightarrow \alpha_{c}  \epsilon^{-1/2}$,
$b^{1/2}_{ab} \rightarrow b^{1/2}_{c} \epsilon^{1/4}$ in $U_1$,
$(\eta_{ab} b_{ab})^{1/2} \rightarrow (\eta_{c} b_{c}^{1/2})\epsilon^{1/4}$ in $U_2$.
The values of the $\alpha$ parameters change as $\alpha_{ab,1}\rightarrow\alpha_{c} \epsilon^{-1/2}$, and $\alpha_{ab,2}\rightarrow\alpha/(\eta\epsilon^{1/2})$.
However, this re-scaling has not been used in this study,
as the ratio between the Fermi velocities $\eta$ is a variable parameter
that can be changed between the two field orientations.

Previous studies on CaKFe$_4$As$_4$ have described the onset upper critical field data up to $60\,$T using the parameters $\lambda_{11}=\lambda_{22}=0$, $\lambda_{12}\lambda_{21}=0.25$, $\eta=0.2$, $\alpha=0.5$ and $\epsilon=1/6$ and the above re-scaling \cite{Meier2016}.
Using these parameters we find that  the onset upper critical field data (open blue and red symbols) clearly fail to describe the full temperature dependence of the upper critical field of CaKFe$_4$As$_4$, as shown in Fig.~\ref{FigSM_TwoBandModels}(a) and (b) (blue and red curves).
As discussed in the main body of text,
we can reasonably describe the upper critical field of CaKFe$_{4}$As$_{4}$
using a two-band model with band parameters for the $s^{\pm}$  and $s^{++}$ pairing (see Table~\ref{CaKFe4As4_TwoBand_Table}),
as shown in Fig.~\ref{FigSM_TwoBandModels}(c) and \ref{FigSM_TwoBandModels}(e), respectively.
We note here that an FFLO state can be stabilized at low temperatures for $H||(ab)$
(Fig.~\ref{FigSM_TwoBandModels}(c) and (e), right axis).

To further test the obtained parameters, we have modelled both the offset and onset critical fields for both field orientations
assuming the $s^{\pm}$ and $s^{++}$ pairing. The parameters for the two cases are listed in Table~\ref{CaKFe4As4_TwoBand_Table}
and the simulations  of the upper critical field are shown in Fig.~\ref{FigSM_TwoBandModels}.
We find that the parameters describing  the offset and onset upper critical fields are similar, with small differences
found for the $\alpha$ values, which are slightly larger as the curvature for $H_{\rm{on}}$ is greater as compared with the $H_{\rm{off}}$ case.
The velocities for $H_{\rm{on}}$ are also slightly smaller due to the larger slopes near $T_{\rm{c}}$,
which scales as $|dH_{\rm{c2}}/dT|_{T_{\rm{c}}}\propto{}1/v^{2}$.
Using the values for $H_{\rm{off}}$ and $H_{\rm{on}}$, we expected a $\sim20\%$ difference in velocities
when $H||c$, and a $\sim10\%$ difference when $H||(ab)$, as shown in  Table~\ref{CaKFe4As4_TwoBand_Table}.
However, independent of the definition of the upper critical field,
the stabilization of an FFLO state in CaKFe$_{4}$As$_{4}$ for $H||(ab)$ is realized in both cases.

\begin{table*}
	\caption{	The different parameters used for the simulations of the upper critical field ( defined either using an offset or onset magnetic field)
using a two-band model and considering different pairing symmetries and magnetic field orientations in CaKFe$_4$As$_4$.}
	\label{CaKFe4As4_TwoBand_Table}
	\begin{tabular}{ccccccccc}
		\hline
		\hline
		& \multicolumn{4}{c}{$\mu_{0}H_{\rm{off}}$} & \multicolumn{4}{c}{$\mu_{0}H_{\rm{on}}$} \\
		& \multicolumn{2}{c}{$s^{\pm}$} & \multicolumn{2}{c}{$s^{++}$} & \multicolumn{2}{c}{$s^{\pm}$} & \multicolumn{2}{c}{$s^{++}$} \\
		%\hline
		& $H||c$ & $H||(ab)$ & $H||c$ & $H||(ab)$ & $H||c$ & $H||(ab)$ & $H||c$ & $H||(ab)$ \\
		\hline
		$\lambda_{11}$ & 0 & 0 & 0.81 & 0.81 &
		0 & 0 & 0.81 & 0.81 \\
		$\lambda_{22}$ & 0 & 0 & 0.29 & 0.29 &
		0 & 0 & 0.29 & 0.29 \\
		$\lambda_{12}$ & 0.5 & 0.5 & 0.1 & 0.1 &
		0.5 & 0.5 & 0.1 & 0.1 \\
		$\alpha_{1}$ & 0.95 & 3.1 & 0.5 & 3.2 &
		1.5 & 4.2 & 1.0 & 3.5 \\
		$\alpha_{2}$ & 0 & 0.7 & 0 & 0 &
		0 & 0 & 0 & 0 \\
		$v_{1}$ ($\mathrm{meV\AA}$) & 510(20) & 253(20) & 380(20) & 195(20) &
		400(20) & 215(20) & 312(20) & 181(20) \\
		$v_{2}$ ($\mathrm{meV\AA}$) & 40(4) & 51(5) & 54(4) & 41(4) &
		36(3) & 43(4) & 44(4) & 38(4) \\
		$\eta$ & $\sim0.006$ & $\sim0.04$ & $\sim0.02$ & $\sim0.04$ &
		$\sim0.008$ & $\sim0.04$ & $\sim0.02$ & $\sim0.045$ \\
		\hline
		\hline
	\end{tabular}
\end{table*}

\end{document}